\documentclass[12pt,preprint]{aastex}

\shorttitle{}
\shortauthors{}

\begin{document}

\title{Time Dependent Nature of the Neptune's Arc Configuration}
\author{K.H. Tsui}
\affil{Instituto de F\'{i}sica - Universidade Federal Fluminense
\\Campus da Praia Vermelha, Av. General Milton Tavares de Souza s/n
\\Gragoat\'{a}, 24.210-346, Niter\'{o}i, Rio de Janeiro, Brasil.}
\email{tsui$@$if.uff.br}
\pagestyle{myheadings}
\baselineskip 18pt

\begin{abstract}

By distinguishing the main arc Fraternite 
 from the minor arcs Egalite (2,1), Liberte, Courage,
 the restricted three-body system
 is extended to a non-conservative restricted four-body system
 with the central body Neptune S, the primary body Galatea X,
 a minor body Fraternite F, and a test body s.
 Through the equations of motion,
 it is shown that the locations
 where the force is null (null points) along the orbit of s
 correspond to the locations of Egalite (2,1), Liberte, and Courage.
 Even if all the arcs were captured by three-body CER sites initially,
 the orbits over the CER potential maxima would be unstable for the minor arcs
 due to the disturbing force of Fraternite with finite mass,
 allowing them to be relocated to the four-body null points. 
 On the other hand, the minor arcs do not have the mass
 to destabilize Fraternite from its three-body CER site,
 which is enlarged from 8.37 degrees to 9.7 degrees
 thus accounting for the Fraternite's span.
 In this restricted four-body system,
 s is under the effects of the potentials of X and F.
 The potential of X drives a long period
 harmonic pendulum oscillation 
 through $\Phi_{sx}=[(n+1)\theta_{s}-n\theta_{x}-\phi_{s}]$
 centered over a four-body null point with $T=1,000\,d$,
 while the potential of F drives a much longer period
 singular pendulum oscillation through $\Delta\theta_{sf}$
 centered over Fraternite with $T_{f}=40,000\,d$.
 In this four-body system, the dynamics of these two oscillations
 generates a time alternating symmetric arc configuration about Fraternite
 with a century long time scale. 
 The non-conservative nature of this system
 could account for the time varying intensity,
 configuration, and disappearance of the minor arcs.

\end{abstract}


\maketitle

\newpage
\section{Time Dependent Arcs of Neptune}

In celestial mechanics, the time varying configuration of the Neptune arcs,
 with Egalite (2,1), Liberte, and Courage,
 extending about $40^{0}$ ahead of the main arc Fraternite,
 \citep{hubbard1986, smith1989}
 stands out as an unsolved problem.
 The first early model, the two-satellite model,
 considered that the arcs were dynamically kept in their locations
 with Galatea doing the radial confinement
 plus a hypothetical Lagrange moon
 doing the azimuthal confinement
 \citep{lissauer1985, sicardy1992}.
 The second was the one-satellite model
 with the inner Galatea and the outer arcs
 in the $86/84$ corotation inclination resonance (CIR)
 with a site of $4.19^{0}$ for azimuthal confinement 
 coupled to the $43/42$ Lindblad resonances (LR) for radial confinement
 \citep{goldreich1986, porco1991, horanyi1993, foryta1996}
 although the main arc Fraternite has a span of about $10^{0}$.
 Later on with more observations
 \citep{sicardy1999, dumas1999},
 and due to the mean motion mismatch and outward radial offset,
 the $43/42$ corotation eccentricity resonance (CER)
 with a resonant site of $8.37^{0}$
 coupled to LR of a massive Fraternite
 to pull on the apsidal line of Galatea
 was proposed to account for the dynamics of the arcs
 \citep{namouni2002}.
 Nevertheless, due to the uncertainties
 of Galatea's eccentricity and the mass of the arc system,
 the CER model still has some open issues,
 in particular, the angular spread of the arcs
 and the irregular spacing among them.
 More intriguingly, the arc intensities are changing in time 
 with some arcs flare up and others fade away
 \citep{sicardy1999, dumas1999}
 and even the arc configuration itself
 appears to be changing in time as well
 with the leading arc Courage appears
 to have leaped over to another CER site
 \citep{depater2005}.
 Most challengingly, the recent HST results
 have indicated that Courage and Liberte apparently have disappeared
 and Egalite (2,1) have reduced in brightness
 \citep{showalter2016}.
 These dynamic properties show that the arcs
 are not in a stable equilibrium configuration.

Before we present our approach,
 let us first recall that the CIR-LR and CER-LR resonance models
 are based on the restricted three-body framework
 with the central body S (Neptune), the primary body X (Galatea),
 and the test body s (arcs).
 In this system, Fraternite and the minor arcs,
 that do not interact with each other,
 are treated on the same basis as test bodies.
 In this three-body system,
 the force acting on s in the SX center of mass reference
 becomes conservative by choosing a frame that rotates with X.
 Consequently, this force can be expressed
 through the gradient of a disturbing potential.
 For s coorbital with X,
 it is well known that this system has three collinear
 and two equilateral triangular Lagrangian points
 that could harbor test bodies around the potential maxima.
 For s corotation with X, we have the CIRs and CERs.
 This disturbing potential of s
 in the rotating frame of X is independent of time
 (implicitly time dependent through the motion of s) 
 and is a function of the parameters of X and s,
 such as inclination and eccentricity.
 As for the two-satellite model,
 the arc system is treated as the superposition
 of two conservative three-body systems.
 
Here, we take a different approach
 by considering a coplanar restricted four-body framework
 \citep{tsui2002}
 that consists of the central body S (Neptune),
 the primary body X (Galatea),
 a minor body F (Fraternite) with a finite mass,
 and a test body s (minor arcs).
 The basic assumption of this model
 is to differentiate the main arc Fraternite
 from the minor arcs,
 by considering specifically additional actions of Fraternite on the minor arcs,
 other than pulling on the apsidal line of Galatea.
 Contrary to the restricted three-body framework,
 the force on s in the fixed inertial frame of space
 is always explicitly time dependent
 due to the presence of X and F.
 Choosing a rotating frame of X
 does not eliminate the time dependence entirely. 
 For this reason, the force is non-conservative,
 and cannot be derived through a potential field
 which makes the potential function approach ineffective.
 Due to the residue eccentricity of X,
 the actions of F could rival or override the actions of X
 to generate the time varying structure of the minor arcs s.
 Even if all the arcs were confined by CER potential initially,
 the minor arcs would be perturbed by Fraternite
 through the time dependent force,
 and energy could be taken from Fraternite
 and delivered to the minor arcs
 to make their orbits around CER maxima unstable,
 while the minor arcs with negligible mass
 are not able to do the same on Fraternite.	
 Directly through the equations of motion of the test body,
 the action of Fraternite has been evaluated
 \citep{tsui2007a, tsui2007b}.
 It is shown that the locations on the Adams ring
 where the force acting on s is null
 are compatible to the locations of Egalite (2,1)
 \citep{tsui2007a}
 and also the different locations occupied by Liberte and Courage over the decades
 \citep{tsui2007b}.
 Here, we follow the approach of Tsui
 but emphasize on the physical fundamentals of the model,
 the mathematical clarity of the development,
 and the interpretation of the results.
 In particular, we present a very long period singular pendulum mechanism
 coupled to a long period harmonic pendulum
 to account for the time dependent nature of the Neptune arcs.

\newpage
\section{Non-Conservative Restricted Four-Body System}

We designate $M,m_{x},m_{f}$ as the masses of the central body S,
 the primary body X, and the minor body F respectively.
 Also
 $\vec{r}_{x}=(r_{x},\theta_{x})$, $\vec{r}_{f}=(r_{f},\theta_{f})$,
 and $\vec{r}_{s}=(r_{s},\theta_{s})$
 are the position vectors of X, F, and s with respect to S.
 Furthermore, $\vec{R}=\vec{r}_{s}-\vec{r}_{x}$
 and $\vec{R'}=\vec{r}_{s}-\vec{r}_{f}$
 are the position vectors of s
 with respect to X and F respectively as in Fig.1.
 Expanding the orbit parameters of the test body s 

\begin{eqnarray}
\nonumber
r_{s}\,
=\,a[1-e\cos(\theta_{s}-\phi_{s})]
\,\,\,,
\\
\nonumber
\theta_{s}\,
=\,\theta_{s0}+2e\sin(\theta_{s0}-\phi_{s})\,
=\,\theta_{s0}+\delta\theta_{s}
\,\,\,,
\end{eqnarray}

\noindent
the equations of motion of s
 in the fixed SX center of mass frame in space are
 \citep{tsui2007a}

\begin{eqnarray}
\nonumber
\frac{d^2r_{s}}{dt^2}\,
 =\,+(\frac{GM}{L})^2(\frac{2e^2}{a})
 -\frac{m_{x}}{M}(\frac{GM}{L})^2
 a^2\frac{1}{R^3}[1-e\cos(\theta_{s}-\phi_{s})]
\\
\nonumber
 +\frac{m_{x}}{M}(\frac{GM}{L})^2
 ar_{x}\frac{1}{R^3}\cos(\Delta\theta_{sx})
 -\frac{m_{x}}{M}(\frac{GM}{L_{x}})^2
 r_{x}^2\frac{1}{r_{x}^3}\cos(\Delta\theta_{sx})
\\
\label{eqno1}
 -\frac{m_{f}}{M}(\frac{GM}{L_{f}})^2
 ar_{f}\frac{1}{R'^3}[1-e\cos(\theta_{s}-\phi_{s})]
 [1-\frac{r_{f}}{a}\cos(\Delta\theta_{sf})]
 \,\,\,,
\\
\nonumber
\frac{1}{r_{s}}\frac{d}{dt}(r_{s}^2\omega_{s})\,
 =\,-\frac{m_{x}}{M}(\frac{GM}{L})^2
 ar_{x}\frac{1}{R^3}
 [\sin(\Delta\theta_{sx})+\cos(\Delta\theta_{sx})\delta\theta_{s}]
\\
\label{eqno2}
 +\frac{m_{x}}{M}(\frac{GM}{L_{x}})^2
 r_{x}^2\frac{1}{r_{x}^3}
 [\sin(\Delta\theta_{sx})+\cos(\Delta\theta_{sx})\delta\theta_{s}]
 -\frac{m_{f}}{M}(\frac{GM}{L_{f}})^2
 r_{f}^2\frac{1}{R'^3}\sin(\Delta\theta_{sf})
 \,\,\,.
\end{eqnarray}

\noindent
Here, $\Delta\theta_{sx,sf}=(\theta_{s}-\theta_{x,f})$,
 whereas $\omega_{s}=d\theta_{s}/dt$ is the angular velocity of s
 about the central body S.
 Furthermore, we also have $L^2=GMa$ with $a$ as the semi-major axis of s,
 $L^2_{x}=GMr_{x}$ and $L^2_{f}=GMr_{f}$.

We note that the $m_{f}$ term in each equation
 is responsible for the non-conservative nature
 of the force acting on s.
 Taking a rotating frame of Galatea X does not do away
 the time dependence of Fraternite F.
 The finite mass of F exerts a force on s in the rotating frame of X.
 Because of libration over a null point,
 the orbit of s in the fixed frame of space is not a simple ellipse,
 but an ellipse superimposed with slow librations.
 Thus, as s returns to the same position in space with respect to X,
 F has slightly displaced with respect to s due to the slow librations,
 making the force on s explicitly time dependent,
 not implicitly time dependent through its movement,
 and therefore the force is non-conservative.
 The standard restricted four-body system
 has two minor bodies interacting strongly initially as a binary system
 under the center of mass frame of S and X.
 On approaching the primary body X, 
 the binary suffers tidal force and gets interrupted
 with one of them captured as a satellite \citep{tsui2002}.
 The present case of Neptune arcs differs from this $(2+2)$ system
 in the sense that the interaction between Fraternite and the minor arcs
 is always very weak comparing to the action of Galatea X.
 Although the action of Galatea is much larger,
 nevertheless it is a fast oscillating periodic force
 which can be removed by taking a long time average on the equations of motion
 to expose the dynamics of the non-conservative part of this complex system.
 In this sense, the non-conservative force is a perturbation
 on the standard three-body system,
 but is large enough to destabilize the three-body CER sites for minor arcs,
 generating the time dependent nature of the arc configuration.
 
Taking a long time average comparing to the orbital period of X,
 the $1/r_{x}^3$ center of mass recoil terms of X go away.
 We note that the $\sin(\Delta\theta_{sx})/R^3$ term in Eq.(2) also goes away
 because the $\sin(\Delta\theta_{sx})$ factor is an odd function
 and the $1/R^3$ factor is an even function
 in the interval of $(-\pi,+\pi)$
 which makes the $\sin(\Delta\theta_{sx})/R^3$ factor an odd function.
 Averaging over one cycle, this term vanishes.
 We then have

\begin{eqnarray}
\nonumber
\frac{d^2r_{s}}{dt^2}\,
 =\,+(\frac{GM}{L})^2(\frac{2e^2}{a})
 -\frac{m_{x}}{M}(\frac{GM}{L})^2
 a^2\frac{1}{R^3}[1-e\cos(\theta_{s}-\phi_{s})]
\\
\nonumber
 +\frac{m_{x}}{M}(\frac{GM}{L})^2
 ar_{x}\frac{1}{R^3}\cos(\Delta\theta_{sx})
\\
\label{eqno3}
 -\frac{m_{f}}{M}(\frac{GM}{L_{f}})^2
 ar_{f}\frac{1}{R'^3}[1-e\cos(\theta_{s}-\phi_{s})]
 [1-\frac{r_{f}}{a}\cos(\Delta\theta_{sf})]
 \,\,\,,
\\
\nonumber
 \frac{1}{r_{s}}\frac{d}{dt}(r_{s}^2\omega_{s})\,
 =\,-\frac{m_{x}}{M}(\frac{GM}{L})^2
 ar_{x}\frac{1}{R^3}
 [\sin(\Delta\theta_{sx})+\cos(\Delta\theta_{sx})\delta\theta_{s}]
\\
\nonumber
 -\frac{m_{f}}{M}(\frac{GM}{L_{f}})^2
 r_{f}^2\frac{1}{R'^3}\sin(\Delta\theta_{sf})\,
\\
\label{eqno4}
 =\,-\frac{m_{x}}{M}(\frac{GM}{L})^2
 ar_{x}\frac{1}{R^3}
 \cos(\Delta\theta_{sx})\delta\theta_{s}
 -\frac{m_{f}}{M}(\frac{GM}{L_{f}})^2
 r_{f}^2\frac{1}{R'^3}\sin(\Delta\theta_{sf})
 \,\,\,.
\end{eqnarray}

\noindent
In these equations, there are two periodic even functions $f(\Delta\theta_{sx})$,
 $f_{1}=1/R^3$ and $f_{2}=\cos(\Delta\theta_{sx})/R^3$,
 with respect to X in the fixed inertial frame of space.
 The cubic inverse distance of R,
 which is a function of $\cos(\Delta\theta_{sx})$ by cosine law,
 can be expanded in harmonics of $\Delta\theta_{sx}$
 through the cosine series
 that reads

\begin{eqnarray}
\nonumber
f\,=\,(b_{0}+f_{n})\,
 =\,b_{0}+2\sum b_{n}\cos(n\Delta\theta_{sx})\,\,\,.
\end{eqnarray}

\noindent
Through the function $f_{n}$,
 and taking also $r_{f}=a$ for coorbital s and F,
 these equations become

\begin{eqnarray}
\nonumber
\frac{d^2r_{s}}{dt^2}\,
 =\,+(\frac{GM}{L})^2(\frac{2e^2}{a})
 -\frac{m_{x}}{M}(\frac{GM}{L})^2\,a^2\,b_{01}
\\
\nonumber
 +\frac{m_{x}}{M}(\frac{GM}{L})^2\,a^2\,eb_{n1}
 \{\cos[(n+1)\theta_{s}-n\theta_{x}-\phi_{s}]
  +\cos[(n-1)\theta_{s}-n\theta_{x}+\phi_{s}]\}
\\
\label{eqno5}
 +\frac{m_{x}}{M}(\frac{GM}{L})^2\,ar_{x}\,b_{02}
 -\frac{m_{f}}{M}(\frac{GM}{L_{f}})^2
 a^2\frac{1}{R'^3}[1-\cos(\Delta\theta_{sf})]
 \,\,\,,
\\
\nonumber
 \frac{1}{r_{s}}\frac{d}{dt}(r_{s}^2\omega_{s})\,
 =\,-\frac{m_{x}}{M}(\frac{GM}{L})^2\,ar_{x}\,2eb_{n2}
\\
\nonumber
 \{\sin[(n+1)\theta_{s}-n\theta_{x}-\phi_{s}]
  +\sin[(n-1)\theta_{s}-n\theta_{x}+\phi_{s}]\}
\\
\label{eqno6}
 -\frac{m_{f}}{M}(\frac{GM}{L_{f}})^2 
 a^2\frac{1}{R'^3}\sin(\Delta\theta_{sf})
 \,\,\,.
\end{eqnarray}

\newpage
\section{Null Points}

In these equations, there are two resonances.
 The first one is the $(n+1)/n$ Xs resonance with X inside F and s,
 and the second one is the $(n-1)/n$ Xs resonance with X outside F and s.
 Keeping only the first resonant term for the Neptune arcs,
 and denoting $\Phi_{sx}=[(n+1)\theta_{s}-n\theta_{x}-\phi_{s}]$,
 we have

\begin{eqnarray}
\nonumber
\frac{d^2r_{s}}{dt^2}\,
 =\,(\frac{GM}{L})^2\frac{1}{a}2e^2
 -\frac{m_{x}}{M}(\frac{GM}{L})^2\,a^2\,b_{01}
 +\frac{m_{x}}{M}(\frac{GM}{L})^2\,a^2\,eb_{n1}\cos(\Phi_{sx})
\\
\label{eqno7}
+\frac{m_{x}}{M}(\frac{GM}{L})^2\,ar_{x}\,b_{02}
 -\frac{m_{f}}{M}(\frac{GM}{L})^2\,
 \frac{a^2}{R'^3}[1-\cos(\Delta\theta_{sf})]\,\,\,,
\\
\label{eqno8}
\frac{1}{r_{s}}\frac{d}{dt}(r_{s}^2\omega_{s})\,
 =\,-\frac{m_{x}}{M}(\frac{GM}{L})^2\,ar_{x}\,2eb_{n2}
 \sin(\Phi_{sx})
 -\frac{m_{f}}{M}(\frac{GM}{L})^2\,
 \frac{a^2}{R'^3}\sin(\Delta\theta_{sf})\,\,\,,
\end{eqnarray}

\noindent
where the coefficients $b_{01}$, $b_{n1}$, $b_{02}$, and $b_{n2}$
 are defined through the Laplace coefficients,
 with $\alpha=r_{x}/a<1$, as

\begin{eqnarray}
\nonumber
 b_{01}\,=\,\frac{1}{2a^3}b^{(0)}_{3/2}(\alpha)\,\,\,,
\\
\nonumber
 b_{n1}\,=\,\frac{1}{2a^3}b^{(n)}_{3/2}(\alpha)\,\,\,,
\\
\nonumber
 b_{02}\,=\,\frac{1}{2a^3}b^{(1)}_{3/2}(\alpha)\,\,\,,
\\
\nonumber
 b_{n2}\,=\,\frac{1}{4a^3}[b^{(n+1)}_{3/2}(\alpha)+b^{(n-1)}_{3/2}(\alpha)]\,\,\,.
\end{eqnarray}

\noindent
In order to establish a relation between
 $\theta_{s}$ of the minor arcs and $\theta_{f}$ of the main arc Fraternite,
 we rewrite the sX variable of the arcs $\Phi_{sx}$
 in terms of the FX variable
 $\Phi_{fx}=[(n+1)\theta_{f}-n\theta_{x}-\phi_{f}]$
 to get $\Phi_{sx}=[\Phi_{fx}+(n+1)\Delta\theta_{sf}-(\phi_{s}-\phi_{f})]$.
 Let us consider a freely orbiting test body s,
 the distance $R'$ between s and F varies periodically in time.
 At some locations of $R'$, the force acting on s could be null.
 By setting the right sides to zero,
 the angular positions of s
 where the force acting on it is null are given by

\begin{eqnarray}
\nonumber
 2e^2+\frac{m_{x}}{M}a^3\,e\,b_{n1}\,
 \cos[\Phi_{fx}+(n+1)\Delta\theta_{sf}-(\phi_{s}-\phi_{f})]
\\
\label{eqno9}
 -\frac{m_{x}}{M}a^3\,[b_{01}-\frac{r_{x}}{a}b_{02}]
 -\frac{m_{f}}{M}\frac{a^3}{R'^3}
 [1-\cos(\Delta\theta_{sf})]\,=\,0
 \,\,\,,
\\
\label{eqno10}
 e\,=\,-\frac{m_{f}}{m_{x}}\frac{a^3}{R'^3}
 \frac{1}{a^2r_{x}\,2b_{n2}}
 \frac{\sin(\Delta\theta_{sf})}
 {\sin[\Phi_{fx}+(n+1)\Delta\theta_{sf}-(\phi_{s}-\phi_{f})]}
 \,\,\,.
\end{eqnarray}

\noindent
In Eq.(9), we keep the first term 
 which is balanced by the third term.
 Substituting the eccentricity of Eq.(10) to Eq.(9),
 considering the center of mass of Fraternite
 be at the maximum of the corotation site
 with $\Phi_{fx}=\pi/2$,
 taking $\phi_{s}=\phi_{f}$,
 writing

\begin{eqnarray}
\label{eqno11} 
\frac{R'}{a}\,
 =\,2\sin(\frac{1}{2}\Delta\theta_{sf})\,\,\,,
\end{eqnarray}

\noindent
and using the following Laplace coefficients for the Neptune system,
the null locations are given by
\citep{tsui2007a}

\begin{eqnarray}
\nonumber 
 2a^{3}b_{01}\,=\,0.26487\times 10^{4}\,\,\,,
\\
\nonumber
 2a^{3}b_{02}\,=\,0.26470\times 10^{4}\,\,\,,
\\
\nonumber
 4a^{3}b_{n2}\,=\,0.39950\times 10^{4}\,\,\,,
\\
\nonumber
 2a^{3}[b_{01}-\alpha b_{02}]\,=\,42.9\,\,\,,
\end{eqnarray}
\\
\begin{eqnarray}
\nonumber 
4\,\frac{\sin^{2}(\Delta\theta_{sf}/2)}{\cos(\Delta\theta_{sf}/2)}
 \cos(n+1)\Delta\theta_{sf}\,
\\
\label{eqno12} 
 =\,-1.5528\times 10^{-4}\frac{m_{f}}{M}(\frac{M}{m_{x}})^{3/2}\,
 =\,-0.5490\times 10^{8}\frac{m_{f}}{M}\,
 =\,-3.5\times 10^{-2}\,\,\,,
\end{eqnarray}

\noindent
where we have taken $M=1.0\times 10^{26}\,Kg$, $m_{x}=2.0\times 10^{18}\,Kg$,
 and $m_{f}=6.4\times 10^{16}\,Kg$.

\newpage
\section{Fraternite Span, Egalite (2,1), Liberte, and Courage}

We have expanded the cubic inverse distance of $R$ through the cosine series 
 to obtain the $\Phi_{sx}$ $(n+1)/n$ corotation resonance variable of Eqs.(7,8)
 which is rearranged to the $\Phi_{fx}$ variable
 to get the null points of Eq.(12),
 assuming the unknown mass of Fraternite be $m_{f}=6.4\times 10^{16}\,Kg$.
 In Eq.(12), the $\cos(n+1)\Delta\theta_{sf}$ factor
 on the left side is a fast oscillating term in space
 that gives $(n+1)$ CER sites along the Adams ring.
 The $\cos(\Delta\theta_{sf}/2)$ and $\sin^{2}(\Delta\theta_{sf}/2)$ factors
 are slow oscillating terms in space
 that modulate the fast oscillating term.
 For small $\Delta\theta_{sf}$,
 the first two factors are most important
 in determining the nearest intercepts.
 The left side of this equation is plotted in Fig.2
 as a function of $\Delta\theta_{sf}$ which shows two minima.
 To understand the first minimum,
 we note that the $\cos(n+1)\Delta\theta_{sf}$ function
 starts out with a central maximum at $(n+1)\Delta\theta_{sf}=0$
 and reaches its first minimum at $(n+1)\Delta\theta_{sf}=\pm\pi$
 on each side forming a complete corotation site of 8.37 degrees
 of unit amplitude with $n=42$.
 However, due to the other factors on the left side,
 the central maximum of cosine is replaced
 by a broad plateau with nearly zero amplitude.
 The numerical solution in Fig.2
 shows that the first minimum has a negative value of about -0.007
 and is slightly shifted outwards to 4.85 degrees on each side
 spanning an angular width of 9.7 degrees.
 This central peak represents only the left side of the equation,
 and it is not the solution of Eq.(12).
 It amounts to the CER site of Fraternite,
 the $\cos(n+1)\Delta\theta_{sf}$ term,
 enlarged by the modulations of the 
 $\sin^{2}(\Delta\theta_{sf}/2)/\cos(\Delta\theta_{sf}/2)$ term
 but with a much smaller amplitude of about 0.01.
 To account for Fraternite span,
 we recall that s is designated for minor arcs only,
 not including Fraternite.
 To overcome this problem,
 we recognize that Fraternite is not a single rigid body
 but a large collection of small masses.
 We can, therefore, regard a given small mass as a test body s itself
 while the center of mass is designated by $m_{f}$
 to validate the application of Eq.(12) for Fraternite span.
 As a result, the whole collection of masses
 would be confined by this enlarged but modified CER potential
 which accounts for the 9.7 degree span of Fraternite.
 As for the second minimum,
 it is located at 12.8 degrees from the center.
 The two intercepts of this minimum with the right side of Eq.(12)
 are at 11.7 and 13.8 degrees
 which correspond to approximately Egalite (2,1) positions
 at 10 and 13 degrees
 \citep{depater2005}.
 Observations showed that its relative intensity to Fraternite
 varied from 17 percent higher in 2002 to 7 percent lower in 2003
 with a 24 percent relative change over a year.
 The angular span of this twin arc Egalite appeared to be 30 percent larger
 in 2005 and 1999 publications than in 1989 Voyager 2 results.
 This widening of Egalite was accompanied
 by a corresponding narrowing of Fraternite
 indicating an exchange of material between them.

Let us now consider the other minor arcs
 by extending the range of $\Delta\theta_{sf}$
 \citep{tsui2007b}.
 The left side of Eq.(12) is plotted in Fig.3 in thick line.
 It shows the fast oscillations of the $\cos(n+1)\Delta\theta_{sf}$.
 These oscillations grow in amplitude
 because of the slow modulations of the $\cos(\Delta\theta_{sf}/2)$
 in the denominator.
 As $\Delta\theta_{sf}$ increases,
 the intercepts are approximately given by
 $\cos(n+1)\Delta\theta_{sf}\,\simeq\,0$
 which are near the zeros of CER sites, not near the maxima,
 separated by $4.19^{0}$.
 Although these intercepts reach to $\Delta\theta_{sf}=\pm\pi$
 symmetrically on both side of Fraternite,
 we have to bear in mind
 that we cannot extend to large values of $\Delta\theta_{sf}$.
 The actions of Fraternite get progressively attenuated over distance
 as such that the arcs can only be confined in its neighborhood.
 This happens to agree with the arc signals
 that attenuate away from Fraternite.
 The minor arcs indeed get less and less intense
 as they get farther and farther away from Fraternite. 
 The roots of Eq.(12), 
 besides ($11.8^{0}$ (Egalite 2), $13.8^{0}$ (Egalite 1)),
 are at
 ($19.3^{0}$, $22.7^{0}$ (Liberte twin 2002)),
 ($27.4^{0}$ (Liberte leading twin 2002), $31.2^{0}$ (Courage 1999)),
 ($35.7^{0}$, $39.7^{0}$ (Courage 2003 resonance jump)),
 ($44.0^{0}$, $48.1^{0}$),
 where the corresponding time varying arcs
 are indicated at the estimated locations
 at the corresponding year.
 The intercepts are grouped in pairs within brackets.
 We have also superimposed a set of CER sites
 with amplitude $0.2$ in dotted line in Fig.3,
 with Fraternite centered at a potential maximum,
 for comparison and discussion purposes.
 As for Liberte, 1999 data showed it was about $3^{0}$
 ahead of its position in Voyager 2 pictures.
 For the 2005 results, the 2002 data appeared to show Liberte as a twin arc
 separated by about $4.5^{0}$ with the leading twin
 at the original Voyager 1989 location,
 while in 2003 it returned again as one single arc at the Voyager location.
 According to Fig.3, this corresponds to locations  
 at $22.7^{0}$ and $27.4^{0}$ separated by $4.7^{0}$.
 With respect to the normally low intensity arc Courage at $31.2^{0}$,
 it flared in intensity to become as bright as Liberte in 1998
 indicating a possible exchange of material between the two arcs.
 Most interestingly, not just the arc intensity is time varying,
 but the arc configuration itself is time varying also.
 Besides Liberte with 1999 data showing a $3^{0}$
 ahead of its position in Voyager 2 results,
 it has been observed that Courage has jumped $8^{0}$ ahead as well
 \citep{depater2005}
 corresponding to a jump from $31.2^{0}$ to $39.7^{0}$ of Fig.3.
 Most puzzling of all is that recently the HST results
 further indicated that Courage and Liberte have disappeared
 while Egalite (2,1) have dimmed in brightness
 \citep{showalter2016}.

\newpage
\section{Dynamics over Null Points}

Through the equations of motion
 in the SX center of mass frame stationary in space,
 we have shown that the locations of null force
 closely match the different locations
 occupied by the minor arcs over the decades.
 While Fraternite is captured by the enlarged CER potential,
 however the minor arcs are unstable over the three-body CERs
 due to the perturbations of Fraternite,
 and have to relocate to the null points.
 Now, let us examine the dynamics of this four-body system
 over the null points 
 by considering the time evolutions
 to understand why the arc configuration changes over time.
 From Eq.(8), the time variation of $\theta_{s}$
 is described by two terms, 
 $m_{x}$ and $m_{f}$ terms, on the right side.
 The $m_{f}$ term is usually a much smaller term,
 except it can be singular as $R'$ approaches zero,
 which amounts to a collision with F.
 However, we should note that F is not a rigid body,
 but a distributed mass with a collection of tiny bodies.
 As s approaches the span of F,
 the test body s would lose its identity
 and merge to F itself.

Considering $r_{s}=a$ and $\omega_{x}=d\theta_{x}/dt$ as constants,
 Eq.(8) can be expressed as

\begin{eqnarray}
\nonumber
\frac{d^{2}}{dt^{2}}\Phi_{sx}\,
 =\,-(n+1)\frac{m_{x}}{M}(\frac{GM}{L})^2\,r_{x}\,2eb_{n2}
 \sin(\Phi_{sx})
 -(n+1)\frac{m_{f}}{M}(\frac{GM}{L})^2\frac{1}{4a^{2}}
 \frac{\cos(\Delta\theta_{sf}/2)}{\sin^{2}(\Delta\theta_{sf}/2)}\,
\\
\label{eqno13}
 =\,-\Omega^{2}\sin(\Phi_{sx})
  -\Omega'^{2}
	\frac{\cos(\Delta\theta_{sf}/2)}{\sin^{2}(\Delta\theta_{sf}/2)}
	\,\,\,,
\end{eqnarray}

\noindent
where the eccentricity $e$ in the first term
 on the right side is given by Eq.(10).
 Making use of the expression of $b_{n2}$,
 the eccentricity reads

\begin{eqnarray}
\nonumber
 e\,=\,-\frac{m_{f}}{m_{x}}\frac{1}{2\times 10^{3}}
 \frac{a^3}{R'^3}
 \frac{\sin(\Delta\theta_{sf})}
 {\cos(n+1)\Delta\theta_{sf}}\,
\\
\label{eqno14}
 =\,-\frac{m_{f}}{m_{x}}\frac{1}{8\times 10^{3}}
 \frac{1}{\sin^{2}(\Delta\theta_{sf}/2)}
 \frac{1}{\cos(n+1)\Delta\theta_{sf}}
 \,\,\,.
\end{eqnarray}

\noindent Considering $\Delta\theta_{sf}=31^{0}$ for Courage,
 and taking the quoted mass of $m_{f}$ and $m_{x}$,
 Eq.(14) gives $e=1.3\times 10^{-4}$.
 On the other hand, the first term on the right side of Eq.(13),
 the $\Phi_{sx}$ term, gives the pendulum oscillation,
 which is driven by X through its mass $m_{x}$
 and is a function of the eccentricity $e$ of the test body.
 This is an oscillation of $(n+1)\theta_{s}$
 with respect to $n\theta_{x}$.
 We recall that in the restrict three-body system,
 the phase angle $\Phi_{sx}$ defines the CER corotation sites.
 As a matter of fact, the variable $\Phi_{sx}$
 is subjected to an arbitrary additive phase $\phi_{site}$
 to label the $(n+1)$ CER sites in the SXs three-body system.
 Here in the present SXFs four-body system,
 the dynamic time evolution of this system shows
 that the phase angle $\Phi_{sx}$ oscillates harmonically
 over a null point labeled by $\phi_{site}$.
 The pendulum frequency is approximately given by

\begin{eqnarray}
\label{eqno15}
\Omega^{2}\,
 =\,(n+1)\frac{m_{x}}{M}2\times 10^{3}\,e\,(\frac{2\pi}{T_{s}})^2\,
 =\,(\frac{1}{2.2\times 10^{3}}\frac{2\pi}{T_{s}})^2\,\,\,.
\end{eqnarray}

\noindent This gives a pendulum period $T=2.2\times 10^{3}T_{s}$
 where $T_{s}$ is the period of the arc.
 Taking $T_{s}=0.45\,d$ gives $T=1,000\,d$.
 We should especially note that this long period of oscillation
 is a direct consequence of the very small eccentricity
 $e=1.3\times 10^{-4}$ of s.

The second term on the right side of Eq.(13)
 is a singular term at $(\Delta\theta_{sf}/2)=0$ with

\begin{eqnarray}
\nonumber
\Omega'^{2}\,
 =\,\frac{(n+1)}{4}\frac{m_{f}}{M}(\frac{2\pi}{T_{s}})^2\,\,\,.
\end{eqnarray}

\noindent
Nevertheless, this second term is not a standard pendulum term.
 To clarify the nature of this $\Omega'^{2}$ term,
 which is independent of the eccentricity $e$,
 we multiply Eq.(13) by $2d\Phi_{sx}/dt$ to get

\begin{eqnarray}
\label{eqno16}
\frac{d}{dt}(\frac{d\Phi_{sx}}{dt})^{2}\,
 =\,2\Omega^{2}\frac{d}{dt}(\cos\Phi_{sx})
  -2\Omega'^{2}\frac{d\Phi_{sx}}{dt}
	\frac{\cos(\Delta\theta_{sf}/2)}{\sin^{2}(\Delta\theta_{sf}/2)}
	\,\,\,.
\end{eqnarray}

\noindent Recalling $\Phi_{sx}=[\Phi_{fx}+(n+1)\Delta\theta_{sf}-(\phi_{s}-\phi_{f})]$
 and considering F be captured by CER-LR with $\Phi_{fx}$ constant,
 we then have

\begin{eqnarray}
\label{eqno17}
\frac{d\Phi_{sx}}{dt}\,
=\,2(n+1)\frac{d}{dt}(\frac{\Delta\theta_{sf}}{2})
\,\,\,,
\end{eqnarray}

\noindent
which allows the second term on the right side
be written as a total derivative.
Consequently, the first integral of the pendulum equation is

\begin{eqnarray}
\label{eqno18}
\frac{1}{2}(\frac{d\Phi_{sx}}{dt})^{2}+[-\Omega^{2}\cos\Phi_{sx}]
 +[-2(n+1)\Omega'^{2}\frac{1}{\sin(\Delta\theta_{sf}/2)}]\,
 =\,C\,\,\,.
\end{eqnarray}

\newpage
\section{Null Point Harmonic Pendulum}

Our analysis of Eq.(8) shows
 that there is a $\Phi_{sx}$ oscillation 
 of $(n+1)\theta_{s}$ with respect to $n\theta_{x}$ over a null point,
 with Fraternite field as a perturbation, indicated by Eq.(18).
 This slow pendulum oscillation comes as a surprise
 in this four-body system.
 But, may be it is not.
 We recall that, in the familiar restricted three-body framework,
 test bodies are captured at the Lagrangian points
 through slow librations around these locations,
 in particular L4 and L5,
 which is a pendulum oscillation
 of $\Phi_{sx}=(1\theta_{s}-1\theta_{x})$
 with $1\theta_{s}$ oscillating with respect to $1\theta_{x}$
 but centered at $\phi_{site}=\pm\pi/3$
 in the $1/1$ coorbital resonance.
 The present case is just the $(n+1)/n$ corotation resonance counterpart.
 To understand qualitatively this complex pendulum oscillation,
 we first consider the $\Phi_{sx}$ oscillation
 with $\Omega'^{2}=0\,(m_{f}=0)$ in the first integral, Eq.(18).
 We note that the inner and outer turning points
 are defined by $(d\Phi_{sx}/dt)^{2}=0$
 where the kinetic energy of s is null.
 The inner/outer turning point
 corresponds to location closer/farther to Fraternite.
 The harmonic potential energy of Galatea X
 is an inverted cosine function with $8.37^{0}$ periodicity,
 due to the negative sign,
 on the lower half plane in a plot against $\Phi_{sx}$
 covering the range $(\Phi_{sx1},\Phi_{sx2})=(-\pi/2,+\pi/2)$.
 This corresponds to a range $(\theta_{s1},\theta_{s2})$
 where $\theta_{s1}/\theta_{s2}$ is the inner/outer turning point
 in terms of $\theta_{s}$.
 The positive kinetic energy term on the upper half plane
 is just the opposite of the potential energy term
 which gives the inner/outer turning point at
 $\Phi_{sx1}/\Phi_{sx2}=(-\pi/2)/(+\pi/2)$.
 The pendulum maximum to maximum amplitude is
 $\delta\Phi_{sx}=(n+1)\delta\theta_{s}=(n+1)(\theta_{s2}-\theta_{s1})=\pi$,
 or $\delta\theta_{s}=4.19^{0}$.
 This case corresponds to a particular value
 of the integration constant $C=C_{0}=0$.
 The first integral allows an arbitrary constant $C_{shift}$ on the right side,
 and it controls the oscillation amplitude.
 This is the same $(n+1)/n$ corotation resonance
 in the three-body system over a CER site with period of $T=2\pi/\Omega$,
 just like over L4 and L5.
 We recall that librations over L4 and L5 are over potential maxima.
 For the present case under the energy conservation picture,
 the librations are over potential minima
 because of the explicit negative sign in front of the potential
 defined in the first integral. 

Now let us consider the effects of the $\Omega'^{2}$ term
 on the harmonic oscillations.
 In the presence of this term, 
 Eq.(18) is the energy conservation
 of the non-conservative four-body system after time averaging.
 The four-body explicit time dependent nature
 of the force acting on s
 is represented by the $(\Delta\theta_{sf}/2)$ variable
 in the $\Omega'^{2}$ term,
 in contrast to the conservative three-body implicit dependence
 of the $\Phi_{sx}$ variable. 
 This term is the potential energy of Fraternite F.
 The first integral, Eq.(18), is the energy conservation law
 of this four-body system. 
 The potential energy is the sum of the F potential,
 which is a monotonic negatively increasing function of $(\Delta\theta_{sf}/2)$
 and becomes singular as $(\Delta\theta_{sf}/2)=0$,
 plus the X potential,
 which is a periodic negative cosine function.
 Since the F potential is a larger term,
 due to the $2(n+1)$ factor, than the X potential, 
 the total four-body potential is a negative singular F potential
 with periodic harmonic ripples of X added on it.
 Along this total potential, we take a negative constant $C$
 such that it intercepts the total potential across a ripple
 bounding a local ripple minimum between two points
 $\theta_{s1}$ and $\theta_{s2}$.
 The corresponding kinetic energy distribution between these two points
 is the positive counterpart
 of the negative ripple potential energy intercepted by $C$.
 Likewise for other ripples with respective negative constants $C$.
 This is the four-body counterpart of the three-body CER libration.
 The harmonic oscillation over this ripple
 is given by the frequency $\Omega$.
 The corresponding radial response
 for the harmonic pendulum oscillations discussed above
 is given by the $\cos\Phi_{sx}$ and the $m_{f}$ terms of Eq.(7)

\begin{eqnarray}
\label{eqno19}
\frac{d^2r_{s}}{dt^2}\,
 =\,\frac{m_{x}}{M}(\frac{GM}{L})^2\,a^2\,eb_{n1}\cos(\Phi_{sx})
 -\frac{m_{f}}{M}(\frac{GM}{L})^2\,
 \frac{1}{4a}\frac{1}{\sin(\Delta\theta_{sf}/2)}\,\,\,.
\end{eqnarray}

\newpage
\section{Fraternite Centered Singular Pendulum}

So far, we have discussed the harmonic oscillations
 centered on a null point in a non-conservative four-body system.
 The potential well in the neighborhood of the null point,
 defined by the integration constant $C$, is harmonic in nature.
 But these oscillations are unable to account for the time dependent arc configuration.
 To resolve this configuration issue, 
 we first note that in the pendulum equation, Eq.(13),
 the $\Omega'^{2}$ term of Fraternite
 is usually a smaller term at large distances
 comparing to the $\Omega^{2}$ term of Galatea,
 but it gets singular as $\sin^{2}(\Delta\theta_{sf}/2)$ approaches zero.
 In the first integral of energy, Eq.(18), 
 the $\Omega'^{2}$ term is multiplied over by a factor of $2(n+1)$,
 which becomes a dominant term in terms of energy as distance gets smaller
 and drives an oscillation centered on Fraternite with a singular potential. 
 This is most evident by writing Eq.(18) as

\begin{eqnarray}
\label{eqno20}
2(n+1)^{2}(\frac{d(\Delta\theta_{sf}/2)}{dt})^{2}+[-\Omega^{2}\cos\Phi_{sx}]
 +[-2(n+1)\Omega'^{2}\frac{1}{\sin(\Delta\theta_{sf}/2)}]\,
 =\,C\,\,\,.
\end{eqnarray}

\noindent 
 These two oscillations, harmonic and singular,
 are superimposed on each other.
 Under this situation,
 the test body s will be accelerated straight to F.
 However, since Fraternite is a distributed mass, not a point mass, 
 s would lose its identity as a test body
 when it reaches within the span of F
 and would be merged with F to become part of Fraternite.
 In this sense, the singularity is rounded off.
 Another test body s will emerge from the other side of F
 by momentum conservation
 after a transit time delay $T_{transit}$
 for the momentum of incoming s to propagate across
 the entire extent of F through collisions.
 We can estimate $T_{transit}$
 by the average time interval between collisions $T_{c}$
 multiplied by the number of collisions $N_{c}$
 needed to cover the longitudinal extent of Fraternite $L_{f}$.
 We, therefore have
 $T_{transit}=T_{c}N_{c}=(2\pi/\nu_{c})(L_{f}/l_{c})$
 where $\nu_{c}$ and $l_{c}$
 are the collision frequency and mean free path respectively.
 In this case, the arcs should reappear
 on the conjugate side of Fraternite
 trailing behind Fraternite.

We have defined the harmonic pendulum frequency $\Omega^{2}$
 through the first integral of the time averaged SXFs four-body system , Eq.(18),
 in terms of the harmonic $\Phi_{sx}$ variable.
 To identify the corresponding frequency $\Omega_{f}$ for oscillations about F,
 we recast Eq.(20) in the standard canonical form
 in terms of $(\Delta\theta_{sf}/2)$ to read

\begin{eqnarray}
\label{eqno21}
\frac{1}{2}(\frac{d(\Delta\theta_{sf}/2)}{dt})^{2}
 +[-\frac{1}{4(n+1)^{2}}\Omega^{2}\cos\Phi_{sx}]
 +[-\Omega_{f}^{2}\frac{1}{\sin(\Delta\theta_{sf}/2)}]\,
 =\,C\,\,\,,
\\
\label{eqno22}
\Omega_{f}^{2}\,
 =\,\frac{\Omega'^{2}}{2(n+1)}\,
 =\,\frac{m_{f}}{8M}(\frac{2\pi}{T_{s}})^{2}\,\,\,.
\end{eqnarray}

\noindent 
While Eq.(18) describes the harmonic oscillation
 with Fraternite field as a perturbation,
 Eq.(21) describes the singular oscillation
 over a much longer time span.
 Both equations amount to Eq.(13) written in different forms,
 and they describe the coupling
 between the harmonic and singular oscillations
 in the non-conservative SXFs four-body system.
 Through $\Omega_{f}$, the period of the singular pendulum is $T_{f}=40,000\,d$
 with $20,000\,d$ on each side of Fraternite
 and $10,000\,d$ each on the forward and backward migrations.
 Furthermore, considering the momentum transit time across Fraternite,
 the total period of the singular oscillation is $(T_{f}+2T_{transit})$.

\newpage
\section{Four-Body Null Point Pendulum Model}

At this stage, it is important to recapture
 the standard restricted SXs three-body system derivation of Lagrangian points.
 The explicit time dependence of X in this system
 is absorbed by choosing a rotating frame of X
 in the SX center of mass system.
 The potential field of this conservative system has locations of maximum,
 which are the null points in the rotating frame of X,
 known as Lagrangian points. 
 The energy of s in this system
 is measured with respect to the rotating frame
 and corresponds to the integration constant C.
 This energy defines the libration
 in the rotating frame about the Lagrangian points.
 In the present SXFs four-body system,
 choosing a rotating frame of X
 does not do away all the explicit time dependencies.
 Nevertheless, the fast orbital time dependence of X can be eliminated
 by taking a time average of the equations of motion
 in the fixed SX center of mass inertial frame.
 In the particular case
 where X and F are in corotation resonance
 and s and F are coorbital,
 the minor arcs s are located at the null points of this system.
 A test particle of a minor arc
 is under a harmonic oscillation
 through $\Phi_{sx}$ over a $\cos\Phi_{sx}$ potential of Eq.(18)
 about a null point,
 and a singular oscillation 
 through $(\Delta\theta_{sf}/2)$ over a $1/\sin(\Delta\theta_{sf}/2)$ potential of Eq.(21)
 about Fraternite. 
 Therefore, the total potential is the sum of the harmonic and singular potentials.
 Thus, the libration over a null point
 is off centered on the harmonic potential maximum
 due to the shift by the singular potential,
 which reflects the non-conservative nature of this four-body system.
 Furthermore, the first integral is derived from the radial equation only,
 and the one-dimensional total potential of oscillation
 is not the general two-dimensional potential field
 of the non-conservative four-body system
 from which the force on s could be derived.
 Such four-body potential field does not exist.

Probably due to the disintegration of a small corotation moon
 with a major part captured by a three-body CER site forming Fraternite F,
 the continuous distribution of tiny bodies along the ring
 represents the extent of the potential energy of F.
 The farther from F the body is,
 the higher would be its potential energy with respect to F.
 These bodies oscillate about F with amplitudes
 set by their initial positions at rest in space with a period $T_{f}=40,000\,d$.
 Along the way to F, as a body s encounters a null point,
 it oscillates over the harmonic potential with $T=1,000\,d$.
 However, because of the non-conservative nature of the force,
 this harmonic oscillation has a finite dwelling time $T_{dwell}$,
 and then s moves down to the next null point, and so on.
 As a result, the test body s would execute
 a singular oscillation with period  $T_{f}$,
 with stop overs $T_{dwell}$ on each null point,
 plus a momentum transit time $T_{transit}$ to traverse Fraternite
 and to emerge on the other side with an equivalent test body s.
 Under this picture, a test body s in a section of the Adam's ring
 dominated by the singular potential of F
 undergoes an extremely slow migration towards F.
 A given null point is therefore visited
 by a time dependent migrating population
 showing a time dependent arc brightness.
 But these null points could dim out
 as the unidirectional migration flux from the ring section comes to an end.
 Through momentum conservation,
 equivalent migration flux would appear on the other side of Fraternite,
 and so would the arcs.

We, therefore, have the following scenario.
 Since the discovery of the Neptune arcs about three decades ago,
 there was a continuous distribution of tiny bodies
 over a section of the Adam's ring
 in the vicinity of Fraternite.
 The distribution was not uniform,
 but with concentrations over the null points.
 Throughout the first decade of discovery,
 the distribution occupied the null points nearest to Fraternite
 and were named as Egalite (2,1), Liberte.
 In the following decade,
 Liberte was seen to split up in twin arcs,
 and furthermore a new arc was discovered named Courage.
 In the recent decade,
 Courage was seen to have leaped forward.
 We believe all these events are the results
 of a very long period forward swing (migration)
 of the singular pendulum oscillation away from Fraternite.
 We recall that the singular pendulum migration
 will distribute the test bodies along the ring
 according to the potential energies they have.
 At a given position $\theta$ from F,
 those bodies with potential energy higher than the potential at $\theta$
 will reach position $\theta$ and go beyond.
 Larger is $\theta$, less flux will pass through there.
 Thus, the brightness of these arcs would decrease with distance.
 Due to the close distance,
 the nearest pair of null points ($11.8^{0}$, $13.8^{0}$)
 which are occupied by Egalite (2,1)
 is continuously visited by migrating fluxes so far.
 Considering the pair ($19.3^{0}$, $22.7^{0}$ (Liberte twin 2002))
 with the outer null point occupied by Liberte twin,
 and the pair ($27.4^{0}$ (Liberte leading twin 2002), $31.2^{0}$)
 with the inner null point occupied by Liberte leading twin,
 they indicate the forward migration through Liberte.
 Furthermore, considering again the pair ($27.4^{0}$, $31.2^{0}$ (Courage 1999)),
 and the pair ($35.7^{0}$, $39.7^{0}$ (Courage 2003)),
 the outer null points also indicate the forward migration through Courage.
 The current disappearance of Liberte and Courage
 might amount to the backward migration towards Fraternite.
 With no more incoming flux from the ring
 to replace the retreating one,
 the distant arcs would dim out one by one.
 Egalite (2,1) would flare up and dim down
 over time in the coming decade
 caused by the retreating non-uniform fluxes,
 that once lit up Liberte and Courage.
 Furthermore after a period of transit time,
 new arcs at the conjugate null points of Egalite (2,1)
 would appear on the other side of Fraternite
 generating a symmetric configuration of arcs about Fraternite
 that alternates in time on a century long timescale.

Through this null point pendulum model,
 we can therefore provide an estimate
 for the mass of Fraternite $m_{f}=6.4\times 10^{16}\,Kg$
 via the locations of the null points,
 and thus the eccentricity of the minor arcs $e=1.3\times 10^{-4}$.
 With this estimate of $m_{f}$,
 it is possible to establish the eccentricity of Galatea 
 through the application of Eq.(5) of
 \citet{namouni2002} to the CER captured Fraternite
 to get

\begin{eqnarray}
\label{eqno23}
e_{x}\,=\,1.4\times 10^{-5}\,\,\,.
\end{eqnarray}

\newpage
\section{Conclusions}

We have followed 
 \citet{namouni2002}
 to take into consideration the mass of Fraternite F
 to separate this main arc F from the minor arcs s,
 not to examine the apsidal pulling on Galatea X,
 but to examine the effects on the minor arcs s
 by using a restricted four-body system
 to complement the CER-LR model
 in fully accounting for the Neptune arcs.
 Due to the interaction between F and s,
 together with Galatea X and the central body Neptune S,
 the force acting on s in the fixed frame of space 
 is non-conservative and therefore not derivable from a potential field.
 This is the critical difference with the restricted three-body system,
 where the force on a test body in the rotating frame of X
 does not depend on time explicitly,
 should we treat all the arcs, F and s, as non-interacting bodies.
 In this three-body system, there are $(n+1)$ CER sites
 on the three-body potential field of s to capture all the arcs,
 similar to the sites of Lagrangian points.
 In the present restricted four-body system,
 the high frequency terms of $\omega_{x}$ are removed
 by taking a long time average in the equations of motion,
 which then allows the analysis of slow interactions.
 Expanding in terms of the orbit parameters of s,
 the arc locations are represented by the null points.
 Oscillations over a null point is described by a harmonic pendulum
 of $(n+1)\theta_{s}$ with respect to $n\theta_{x}$
 with a period $T=1,000\,d$.
 However, due to the actions of F,
 this harmonic oscillation is unstable
 which allows s to leave a null point eventually
 to follow the singular pendulum oscillation centered over Fraternite
 with a much longer period $T_{f}=40,000\,d$,
 generating a symmetric arc configuration that alternates in time.
 On the other hand, with negligible mass,
 the minor arcs are unable to destabilize Fraternite
 over its three-body CER site.
 Since these null points are with respect to Fraternite,
 and Fraternite is captured by the $43/42$ CER-LR,
 thus the entire arc system follows the CER-LR mean motion.
 By considering Fraternite F as an aggregate of small masses,
 each mass can be treated as a test body s also.
 This generates modulations on the $43/42$
 corotation resonance potential of X,
 which enlarges the CER site of Fraternite to 9.7 degrees
 that accounts for the Fraternite span.
 

\clearpage
\begin{figure}
\plotone{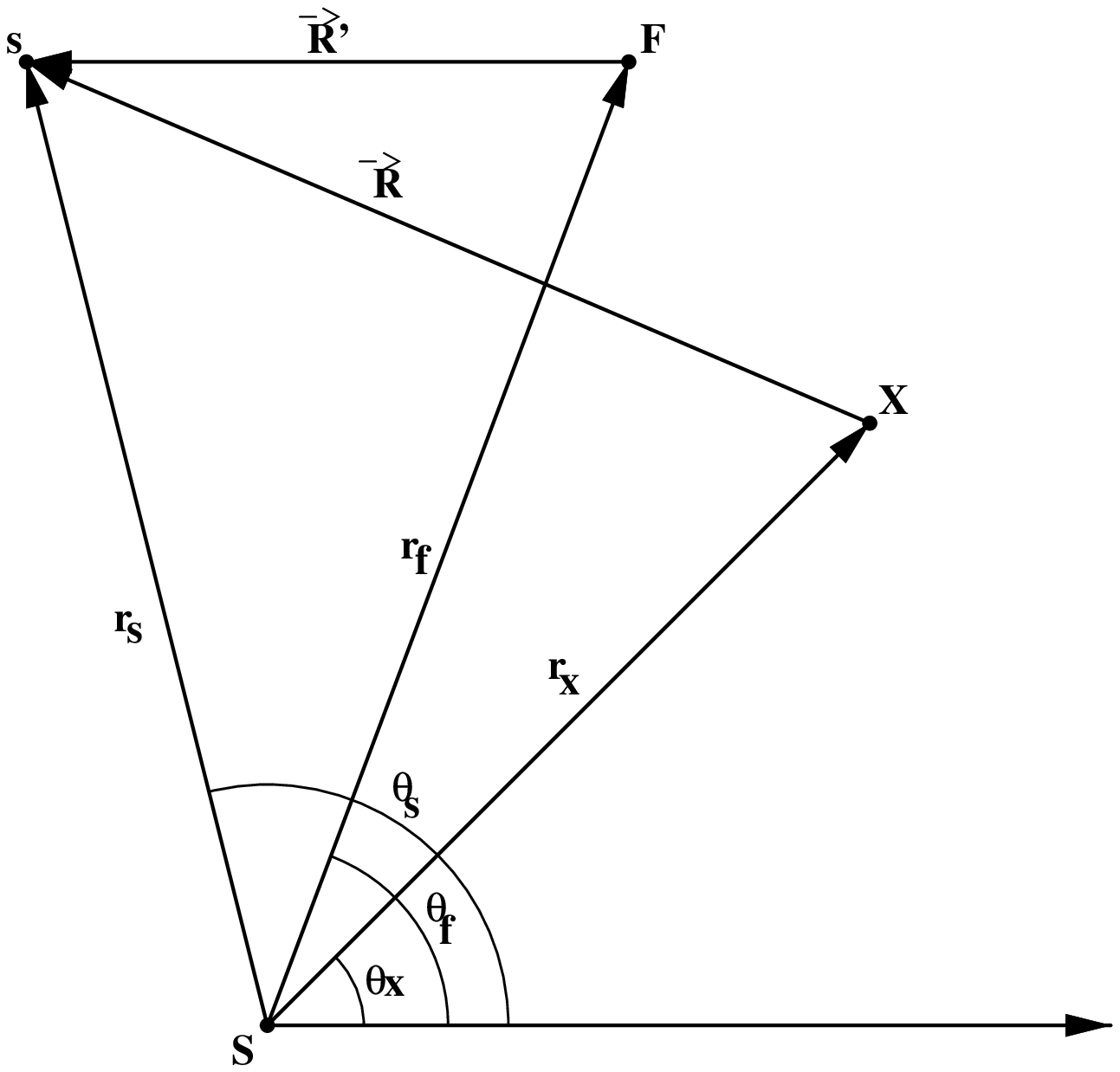}
\caption{This shows the configuration of the coplanar S-X-F-s
 four-body system where S is the central body,
 X is the primary body, F is a minor body, and s is a test body.
 The longitudes are denoted by $\theta$ with appropriate subscripts.}
\label{fig.1}
\end{figure}

\clearpage
\begin{figure}
\plotone{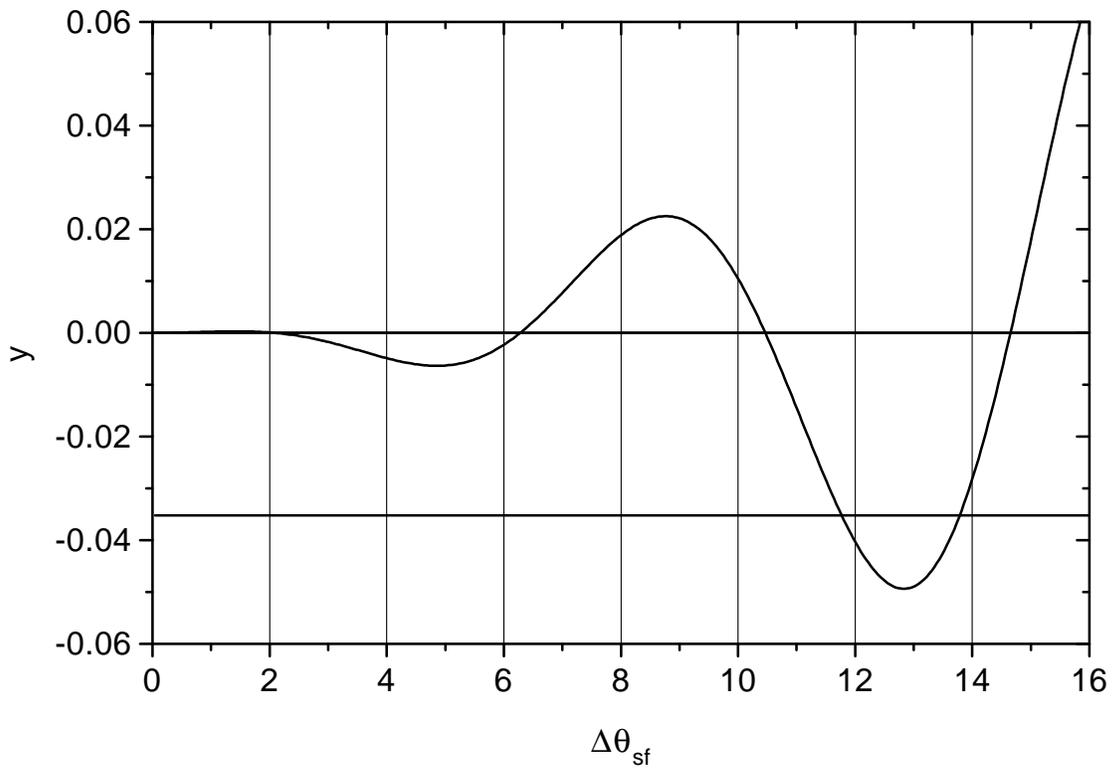}
\caption{The left side of Eq.(12), denoted by the y label,
 is plotted as a function of $\Delta\theta_{sf}$ in degree.
 The right side is a constant and is represented by a horizontal line.
 The intercepts give the roots that define the locations
 where the time averaged force is null.}
\label{fig.2}
\end{figure}

\clearpage
\begin{figure}
\plotone{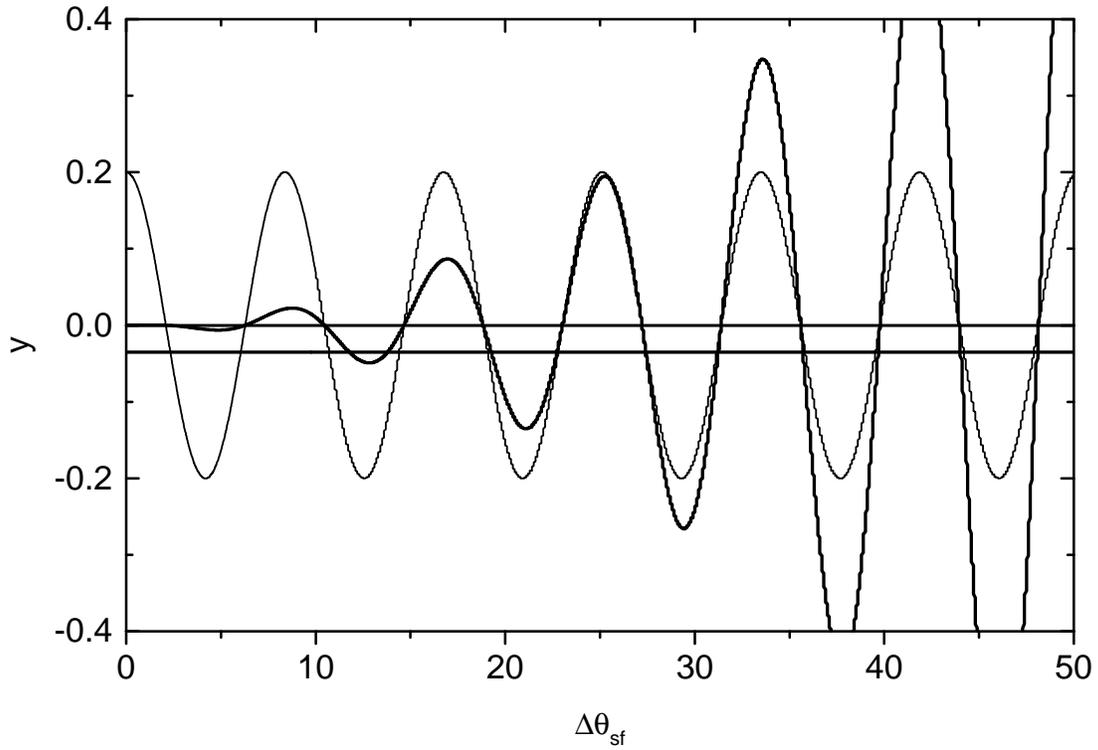}
\caption{The left side of Eq.(12)
 is plotted as a function of $\Delta\theta_{sf}$ in degree
 over a larger range
 with the intercepts covering all the minor arcs.
 The CER sites with Fraternite centered at a potential maximum
 with amplitude $0.2$
 are also shown in dotted line for comparisons.}
\label{fig.3}
\end{figure}

\end{document}